\documentclass[10pt,twocolumn]{article}
\usepackage{amstex}
\usepackage{amssymb}
\begin{document}
\newcommand{\half}{\frac{1}{2}}
\newcommand{\IZ}[1]{\bar{#1}}
\newcommand{\Deta}{\eta^{\dagger}}
\newcommand\Bpsi{\boldsymbol{\psi}}
\title{
\begin{flushright}
  \small UMP-97/65
\end{flushright}
\vskip5mm
\large\bf
The exact realisation of the Lanczos Method for a quantum Many-Body System}
\author
{\large N.S. Witte\footnote{E-mail: {\tt nsw\@physics.unimelb.edu.au}}\\
{\it Research Centre for High Energy Physics,}\\
{\it School of Physics, University of Melbourne,}\\
{\it Parkville, Victoria 3052, AUSTRALIA.} }
\maketitle
\begin{abstract}
The Lanczos process has been analytically and exactly carried out for
the spin 1/2 isotropic XY chain in the thermodynamic limit, yielding a
form for the Lanczos coefficient $ \beta^2(s) $. This coefficient has
a monotonic variation for real positive $s$ and confirms a general
theorem on the ground state properties of extensive Many-body Systems.
The Taylor expansion of the coefficient about $ s=0 $ has a finite
radius of convergence, and ground state estimates based on a finite
truncation of this are shown to be asymptotic.
\end{abstract}

The Lanczos Algorithm is one of the few reliable general methods for 
computing
ground state and some excited state properties of interacting quantum
Many-body Systems. It has been traditionally employed as a numerical
technique on small finite systems, with attendant round-off error
problems, although the main obstacle to its further development is the
rapid growth of the number of basis states with system size. We
wish to demonstrate that the Lanczos Method can be used exactly and
analytically for quantum Many-body Systems, and in addition where
the thermodynamic limit can also be treated exactly. We demonstrate this
for a member of an exactly solvable class of such systems, those
falling into the universality class of the two-dimensional Ising Model,
namely the isotropic spin 1/2 XY Model. 
In fact we have provided another notion of solvability, both
in the mechanics of constructing the solution and in terms of its
general applicability.
In addition we state some general theorems arising from the Analytic
Lanczos Method Formalism, which apply to both integrable and non-integrable
models, where in the latter case a truncation in an expansion 
necessarily takes place. 
Given the exact results for our model the effects
of this truncation can be examined in minute detail, and we present
such a convergence study.

In our convention the Lanczos process generates an orthonormal basis
$ |\psi_n\rangle $ from a suitable trial state $ |\psi_0\rangle $
with the recurrence\cite{eigen-S,eigen-P}
\begin{equation}
  H|\psi_n\rangle = \beta_n|\psi_{n-1}\rangle
                  + \alpha_n|\psi_{n}\rangle
                  + \beta_{n+1}|\psi_{n+1}\rangle \ .
\end{equation}
All of our attention focuses on the Lanczos coefficients so defined,
$ \alpha_n $ and $ \beta^2_n $, rather than the basis states and we
perform these steps exactly.
It was found recently\cite{pexp-exactgse-HW,alm-mops-W} that the 
coefficients arising from an extensive
Many-body System satisfy a confluence property 
$ \alpha_n(N)/N \to \alpha(s) $ and 
$ \beta^2_n(N)/N^2 \to \beta^2(s) $ as $ n,N \to \infty $ 
where the scaled Lanczos iteration number is $ s=n/N $.
With such a description, and employing theorems on the extremal zeros of
Orthogonal Polynomials, 
a number of exact general theorems\cite{alm-W} for the
ground state energy\cite{pexp-exactgse-HW}, 
ground state averages\cite{pexp-2dhm-WHW} and 
gaps\cite{pexp-tgap-HWW,alm-mops-W} of an arbitrary extensive Many-body 
System were found.  For example the ground state energy density is
given by
\begin{equation}
    \epsilon_0 = \inf_{s>0} [\alpha(s)-2\beta(s)] \ .
\label{gse-theorem}
\end{equation}

For the $ N $ site chain, our model Hamiltonian is taken to be
\begin{equation}
 H = \sum^{N}_{i=1} \left[ S^x_iS^x_{i+1}+S^y_iS^y_{i+1} \right] \ ,
\label{xy-ham}
\end{equation}
where the label $ i $ index the sites, and where we only consider
the c-cyclic problem, as the thermodynamic limit will be taken.
This model, being at the isotropic point, is at a critical point of
an anisotropic family of models, and consequently has vanishing 
excited state gaps.
The trial state was taken to be the classical N\'eel state
even though this state is in fact,  as far from the true
ground state as one could possibly be but its simplicity allows
for ease of moment computation, certainly compared to alternatives.
So as a ``natural'' choice we proceed by only considering this 
choice.
In the spin picture the N\'eel state $ |{\cal N}\rangle $ will be 
represented by
$ | \uparrow \downarrow \uparrow \downarrow \ldots 
    \uparrow \downarrow \uparrow \downarrow \rangle $, where the
quantisation axis is in the $z$-direction.
If we denote the vacuum state for the quasiparticles by 
$ |0\rangle $ then the N\'eel state can be expressed in terms of 
the quasiparticle operators $ \Deta_q $ by
\begin{equation}
  |{\cal N}\rangle
    = \prod_{q\in Z} {1+\Deta_q\Deta_{\IZ{q}} \over \sqrt{2}}
               |0\rangle \ .
\label{Neelrep}
\end{equation}
Here the spectrum of allowed momentum values have been divided up into 
the set $ Z \equiv [-\pi, -\pi/2) \cup (0,\pi/2) $
and its complement $ \bar{Z} $, and there is a bijection from $ Z $ to
$ \bar{Z} $ defined by
$ \bar{q} \equiv -q\!-\!\pi $ if $ q \leq 0 $ and 
$ \bar{q} \equiv -q\!+\!\pi $ if $ q  > 0 $.
Basically the N\'eel state is created from the quasiparticles vacuum state
by pairs of excitations whose momenta sum to $ \pm\pi $.

One first constructs the cumulant generating function
from its Taylor series expansion about $ t=0 $,
\begin{equation}
   \langle {\cal N}| e^{tH} |{\cal N} \rangle = 
   \exp\left\{ \sum^{\infty}_{n=1} \nu_n {t^n \over n!} \right\} 
   \equiv e^{F(t)} \ ,
\label{def-F}
\end{equation}
and the exact generating function $ F(t) $ was found to be\cite{texp-W-97}
\begin{equation}
  F(t) = {N \over \pi} 
           \int^{\pi/2}_{0}\, dq \ln\cosh(t\Lambda_q) \ ,
\label{ft-result}
\end{equation}
where the quasiparticle energy is $ \Lambda_q = |\cos(q)| $.
So from the Taylor series expansion about $ t=0 $ we have the cumulants
\begin{alignat}{2}
  \nu_{2n}
    = {2^{2n}\!-\!1 \over 4n}{(2n)! \over (n!)^2} B_{2n} N
    \qquad n \ge 1 \ ,
\label{cumulant}
\end{alignat}
where $ B_{2n} $ are the standard Bernoulli numbers.

Our moment generating function is also essentially the characteristic
function of the underlying measure defining the Lanczos Process, so
a Fourier inversion integral yields the weight function 
$ w(\epsilon) $ in which the only parameter is the system size $ N $.
In what follows we will only need the leading order contribution as
$ N \to \infty $ so we perform a stationary phase approximation yielding
\begin{equation}
   \ln w(\epsilon) = 
   N \left\{ \xi \epsilon + 
             {1\over \pi}\int^{\pi/2}_0 dq \ln\cosh(\xi\Lambda_q)
     \right\} + {\rm O}(1) \ ,
\end{equation}
where $ \xi(\epsilon) $ is the stationary point determined from  
\begin{equation}
  \epsilon = -{1\over \pi}\int^{\pi/2}_0 dq\,\Lambda_q \tanh(\xi\Lambda_q) \ ,
\label{et-result}
\end{equation}
and $ \epsilon = E/N $ the energy density.

A key link to complete our analysis is a very recent result
from the theory of random Matrix Ensembles\cite{matrix-CI-97} in 
which the Lanczos coefficients are determined from the weight function to 
leading order in $ n,N \to \infty $ by the supplementary condition
\begin{equation}
  0 = \int^{\alpha+2\beta}_{\alpha-2\beta} d\epsilon\, 
      { u'(\epsilon) \over \sqrt{4\beta^2-(\epsilon\!-\!\alpha)^2} } \ , 
\end{equation}
and the normalisation condition
\begin{equation}
  n = {1\over 2\pi}\int^{\alpha+2\beta}_{\alpha-2\beta} d\epsilon\, 
  { \epsilon u'(\epsilon) \over \sqrt{4\beta^2-(\epsilon\!-\!\alpha)^2} } \ , 
\end{equation}
where $ u(\epsilon) = -\ln w(\epsilon) $.
This pair of equations arises from the continuum Coulomb gas approach to
the eigenvalue distribution of random Matrix Ensembles, in the solution
for the density of eigenvalues $ \sigma(\epsilon) $, where the
$ u(\epsilon) $ plays the role of a confining one-body potential. 
The explicit solution for this charge density can be constructed, but we do 
not require this here. 
An equivalent derivation\cite{alm-continuum-W} of the above results, which does
not involve Matrix Ensembles, employs a continuum Coulomb gas approach to the 
distribution of stationary points arising from the leading order 
stationary phase approximation to the Selberg integral expression for the
Hankel determinants.

In our model $ \alpha(s)=0 $ and our fundamental result then is the following
pair of implicit equations for $ \beta^2(s) $
\begin{equation}
  \begin{split}
   s & = -{1\over \pi}\int^{2\beta}_{0}d\epsilon\,
          {\epsilon\xi \over \sqrt{4\beta^2-\epsilon^2}}
     \\
   \epsilon & = -{1\over \pi}\int^{\pi/2}_{0}dq\,
           \cos(q)\tanh(\xi\cos(q)) \ .
  \end{split}
\label{exact-xy-alm}
\end{equation}

\begin{figure}[htb]
 \vskip 20.0cm
 \includegraphics{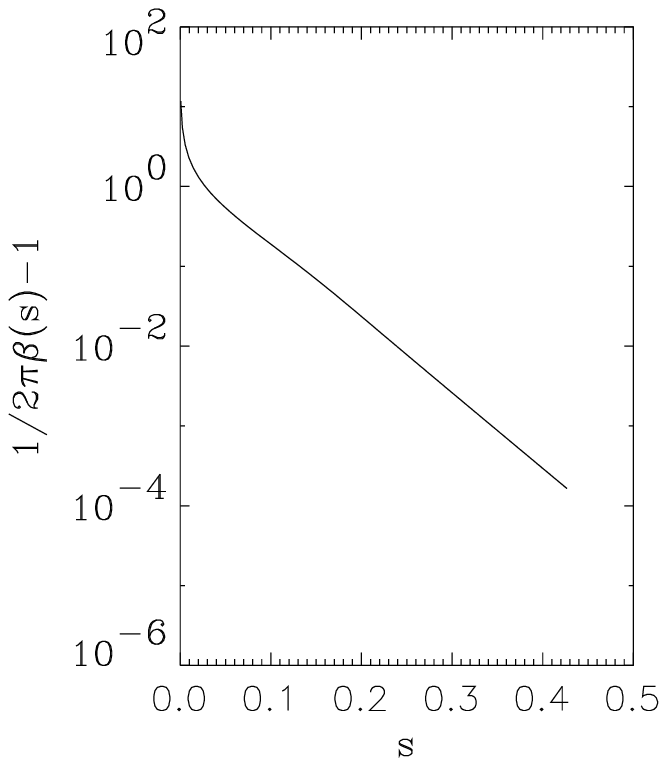}
 \caption
   [Dependence of the Lanczos coefficient $ \beta(s) $ on the 
    scaled Lanczos iteration number $ s=n/N $.]{}
\label{beta}
\end{figure}

In Figure~\ref{beta} we have plotted the dependence of $ \beta $ on 
$ s $ in a form which illustrates the ground state theorem,
Eq.(\ref{gse-theorem}), ($ \epsilon_0 = -1/\pi $). 
Clearly $ -2\beta(s) $ approaches $ \epsilon_0 $ as $ s \to \infty $,
i.e. there is no minima but an infimum, and the monotonicity of this
Lanczos coefficient can be easily shown from the defining equations
\begin{equation}
  {ds \over d\beta^2} = {2\over \pi}\int^{\xi_0}_{0}
                        {d\xi \over \sqrt{4\beta^2-\epsilon^2}} \ ,
\end{equation}
with $ 2\beta = \epsilon(\xi_0) $.
The asymptotic behaviour of this Lanczos coefficient as 
$ s \to \infty $ is also demonstrated in Figure~\ref{beta}, which 
can be analytically shown to be
\begin{equation}
   1-2\pi\beta(s) \sim \delta_{\infty}\exp(-4\pi\sqrt{3}\;s) \ ,
\label{beta_asymp}
\end{equation}
where the pure number $ \delta_{\infty} \approx 1.77799 $.
This is found using the leading order solution for $ \xi(\epsilon) $ as 
$ \xi \to \infty $
\begin{equation}
  \xi \sim {\pi\over 2\sqrt{6}} 
           \left\{  \left(1+{\epsilon\over \epsilon_0}\right)^{-1/2}
                   -\left(1-{\epsilon\over \epsilon_0}\right)^{-1/2}
           \right\} \ .
\label{large-xi}
\end{equation}

\begin{figure}[htb]
 \vskip 20.0cm
 \includegraphics{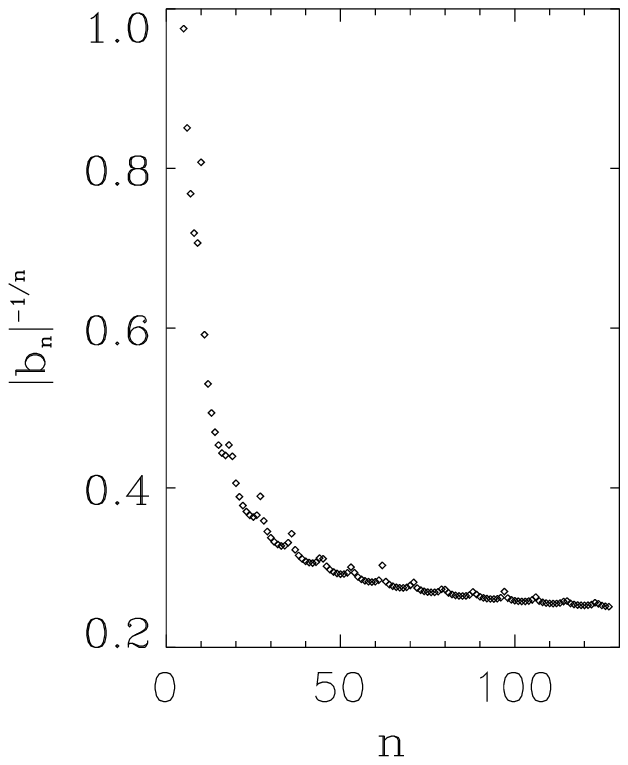}
 \caption
   [The nth root of the Taylor series coefficient $ b_n $ defined
    through the Lanczos coefficient $ \beta^2(s) = \sum_{n=1}b_ns^n $
    versus index $ n $, indicating the radius of convergence of this
    series.]{}
\label{coeff}
\end{figure}

\begin{figure}[htb]
 \vskip 20.0cm
 \includegraphics{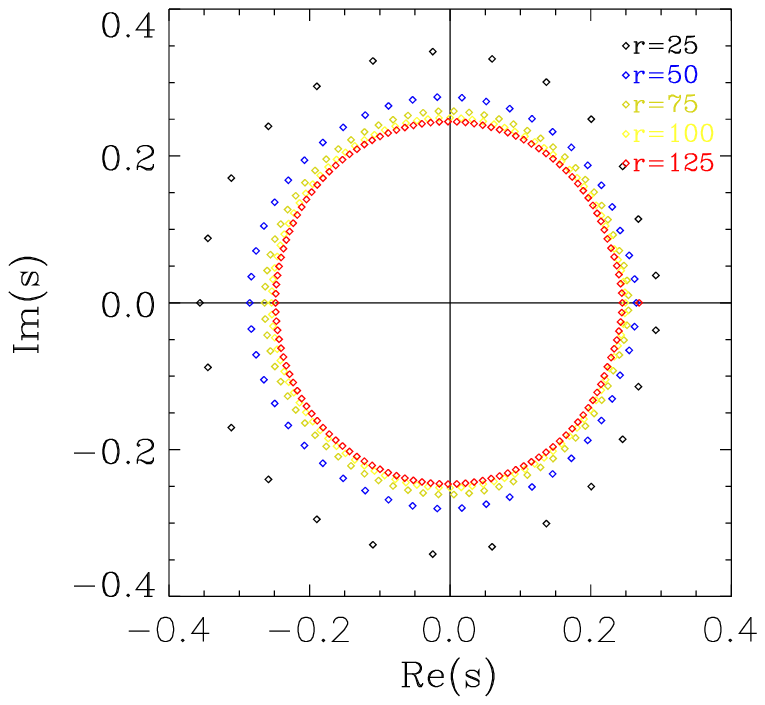}
 \caption
   [The complex zeros of the truncated Taylor series expansion for
    $ \beta^2(s) $ in the complex $ s $-plane, for truncation orders
    $ r= $25, 50, 75, 100, and 125.]{}
\label{zeros}
\end{figure}

Of great interest is the Taylor expansion of $ \beta^2(s) $ about 
$ s=0 $\cite{pexp-L89,pexp-1st-H}, the ''Plaquette Expansion'',
as this forms the basis of the application of the general theorems to
non-integrable models. Using the exact defining equations one can perform
the two reversions of series developments, which has the initial terms
\begin{equation}
\begin{split}
 \beta^2(s) =
 &
	 {\frac {1}{2^{2}}}s
	-{\frac {3} {2^{2}}}s^{2}
	+{\frac {1}{3}}s^{3}
	+{\frac {5}{3^{2}}}s^{4}
	+{\frac {17}{3.5}}s^{5}
 \\
 &
	+{\frac {2.7.127}{3^{3}.5^{2}}}s^{6}
	+{\frac {2^{2}.61.367}{3^{4}.5^{2}.7}}s^{7}
	+{\frac {5.137}{7^{2}}}s^{8}
 \\
 &
	+{\frac {2.17.119417}{3^{6}.5.7^{2}}}s^{9}
	-{\frac {11.103.299539}{3^{8}.5^{3}.7^{2}}}s^{10}
        +\ldots
\end{split}
\label{pe-xy}
\end{equation}
This coincides exactly with an early calculation\cite{pexp-1d-H} and
an independent one using the cumulants directly, although we have 
calculated many more terms.
To indicate the asymptotic behaviour of the coefficients in this Taylor 
expansion, and the convergence of the series we have plotted the nth
root of the coefficients in Figure~\ref{coeff} versus the index. There
is a clear finite limiting value indicating a finite radius of convergence
for the expansion, and an interesting quasi-periodic cusp structure
superimposed.
Another indication of this convergence is given in Figure~\ref{zeros}
where the complex zeros of the truncated $ \beta^2(s) $ are plotted in the
complex $s$-plane for a sequence of truncation orders. 
Clearly here the distribution of zeros
approaches a limiting curve, with a simple shape about $ s=0 $ and at a 
finite distance from the origin. 
As the Lanczos Process terminates when
$ \beta_n \to 0 $ at some nonzero $ n $, this curve gives the breakdown of
the Plaquette Expansion (the exact $ \beta^2(s) $ for real $ s>0 $ is nonzero). 
One can in fact prove this exactly from considerations of the analytic 
behaviour of $ \epsilon(\xi) $ as a function of complex $ \xi $.
This transformation of the $ \xi $-plane is analytic except on the
imaginary axis $ |\Im\,\xi| > \pi/2 $ with a sequence of branch points
at $ \xi_{k} = i(k+\half)\pi $. The real part of $ \epsilon $ has a
discontinuity across these cuts (the defined value on the cut is zero),
and both parts have inverse square-root singularities when approaching
the branch points from one direction along the imaginary axis.
The closest zero of $ \epsilon(\xi) $ to the origin is relevant in
determining the radius of convergence and this occurs as 
$ \xi \to \xi^{+}_{0} $ (in fact all the zeros lie on the imaginary axis,
some at $ \xi^{+}_{k} $) and therefore the value of $ s_0 $ is given by
\begin{equation}
   s_0 = {1\over \pi} \left|
         \int^{\pi/2}_{0} d\sigma\, \Im\,\epsilon(\sigma) \right| \ ,
\label{radius-def}
\end{equation}
where $ \sigma = \Im\,\xi $. The integrand has the following series
development
\begin{equation}
  \Im\,\epsilon(\sigma) = {k_0\!+\!1\over \sigma}
  + {1\over \sigma}\sum^{\infty}_{k=k_0+1}
            \left[ 1-{(k\!+\!\half)\pi \over  
                   \sqrt{(k\!+\!\half)^2\pi^2-\sigma^2}}
            \right] \ ,
\label{imag-energy}
\end{equation}
where $ k_0 = \lfloor \sigma/\pi-\half \rfloor $ .
Using this it is easy to show that
\begin{equation}
\begin{split}
  s_0 & = -{2\over \pi}\log \prod^{\infty}_{odd\: l>0}
        \left[ \half\sqrt{1\!-\!{1\over l}}+\half\sqrt{1\!+\!{1\over l}}
        \right]
      \\
      & \approx
        0.2396967390143423611761660826222 \ .
\end{split}
\label{radius-conv}
\end{equation}

\begin{figure}[htb]
 \vskip 20.0cm
 \includegraphics{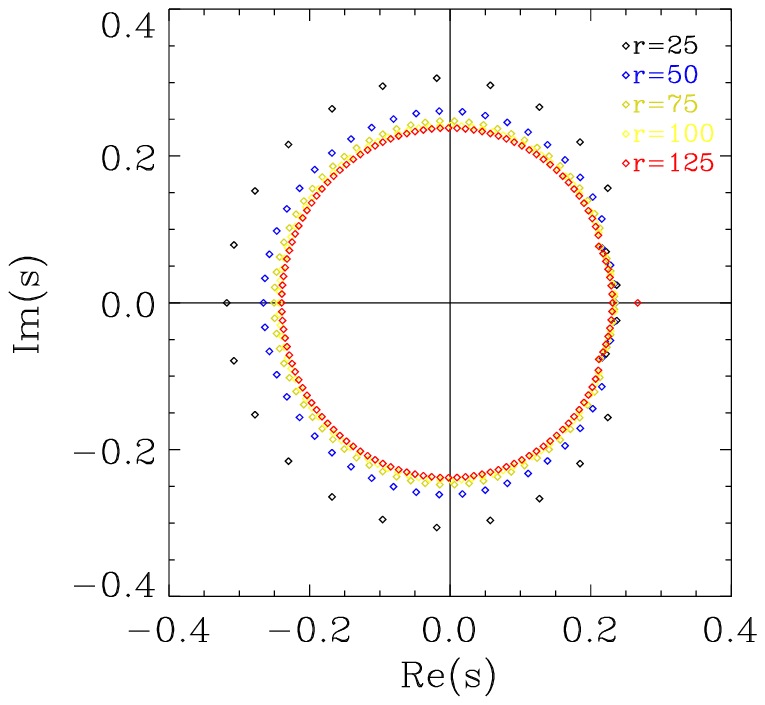}
 \caption
   [The complex zeros of the truncated Taylor series expansion for
    $ d\beta^2(s)/ds $ in the complex $ s $-plane, for truncation orders
    $ r= $25, 50, 75, 100, and 125.]{}
\label{deriv}
\end{figure}

\begin{figure}[htb]
 \vskip 20.0cm
 \includegraphics{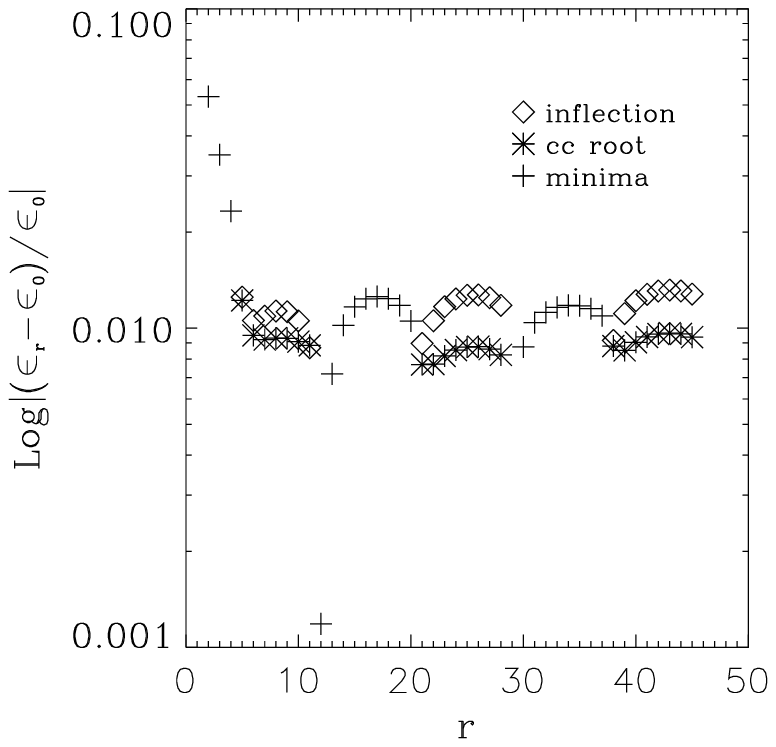}
 \caption
   [The relative error in estimates of the ground state energy using
    the exact ground state energy theorem and the truncated Taylor
    series expansion for $ \beta^2(s) $, versus the truncation order
    $ r $.]{}
\label{error}
\end{figure}

Another relevant set of zeros are those of 
$ d\beta^2(s)/ds $ and these are plotted in Figure~\ref{deriv} for the same
sequence of truncations. At finite truncations a finite minima or a complex
conjugate pair of stationary points (i.e. an inflection point) of $ \beta^2(s) $
always exists (this pair lie close to the positive real axis),
and these can be seen to lie just inside the curve where the Lanczos
process breaks down.
Finally we plot the relative error in the ground state energy estimates
in Figure~\ref{error} using the truncated series directly in the theorem,
Eq.~(\ref{gse-theorem}), versus truncation order.
After an initial reduction in the error to the one percent level, the error
just continues to fluctuate about this level.
The reason why no further improvement is possible with a direct application
of the ground state theorem is that the minima (or inflection point) is
pinned just inside the radius of convergence, whereas it should tend to
infinity in order to converge.

We have found the exact solution to the Lanczos process for a quantum
Many-body System in the thermodynamic limit using results from
the theory of random Matrix Ensembles and Orthogonal Polynomial Systems.
Our exact method applies to any model where the cumulant generating function
can be found, where it should be noted that in this case it was not necessary 
to fully diagonalise the problem in order to find this. 
Our exact solution provides an illustration of some general theorems applying 
to arbitrary extensive Many-body Systems and also demonstrate that these
theorems continue to work as an approximate technique
even at a critical point in the model, albeit to only a semi-quantitative level.
The analytic Lanczos method has recently been applied to a related model with
a quantum phase transition, the Ising model in a transverse 
field\cite{alm-itf-WS,alm-itf-WH}, to elucidate the dependence of the Lanczos
Process away from, near and at criticality.
This work was supported by the Australian Research Council.

\bibliographystyle{aip}
\bibliography{xy,texp,pexp,alm,moment}

\listoffigures
\vfill\eject

\eject

\eject

\eject

\eject

\eject

\end{document}